 \documentclass[12pt,preprint,aps,showpacs]{revtex4}  
 \usepackage[latin1]{inputenc}
 \usepackage{amsmath}
 \usepackage{epsfig}
 \usepackage{subfigure}
 \usepackage{placeins}
 \usepackage{floatflt}
 \usepackage{amsmath}
 \usepackage{amssymb}
 \usepackage{wrapfig}
 \usepackage[]{tocbibind}
 \usepackage{graphicx}

 \usepackage{ifthen}
 \usepackage{times}

\makeatletter
\newcommand\figcaption{\def\@captype{figure}\caption}
\newcommand\tabcaption{\def\@captype{table}\caption}
\makeatother

\newlength{\mylinewidth}
\begin{document}

\title{Search for ppK$^-$ with proton induced reactions at GSI}

\author{L. Fabbietti for the FOPI collaboration}

\address{Excellence Cluster Universe, Technical University Munich,\\
Garching, 85748, Munich\\
$^*$E-mail: laura.fabbietti@ph.tum.de\\
www.universe-cluster.de}

\begin{abstract}
Several experiments engaging kaon beams have been carried out to study the existence of deeply bound kaonic state. We proposed an alternative method, which exploits  a p+p reaction at 3.5GeV incoming energy to search for the most elementary of the kaonic states: ppK$^-$.
This experiment is planned with  the FOPI spectrometer at GSI and a silicon- based trigger system has been developed to increase the detection capability of the existing system for the reaction of interest.
Several tests have been performed under realistic beam  conditions to test the feasibility of the measurement and the proof of principle of the trigger system. The positive results of such measurements are shown in this contribution together with the expected rates extracted from simulations.

\end{abstract}
\maketitle

\section{Introduction}\label{sec:0}
\begin{figure}[h]
\begin{minipage}[b]{5.5cm}
\centering
\includegraphics*[width=5cm]{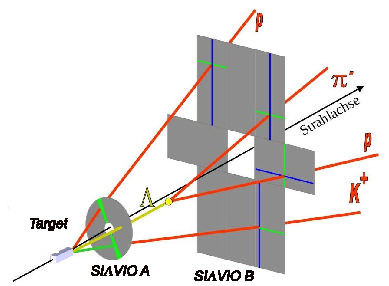}
\end{minipage}
\begin{minipage}[b]{5.5cm}
\centering
\includegraphics*[width=5cm]{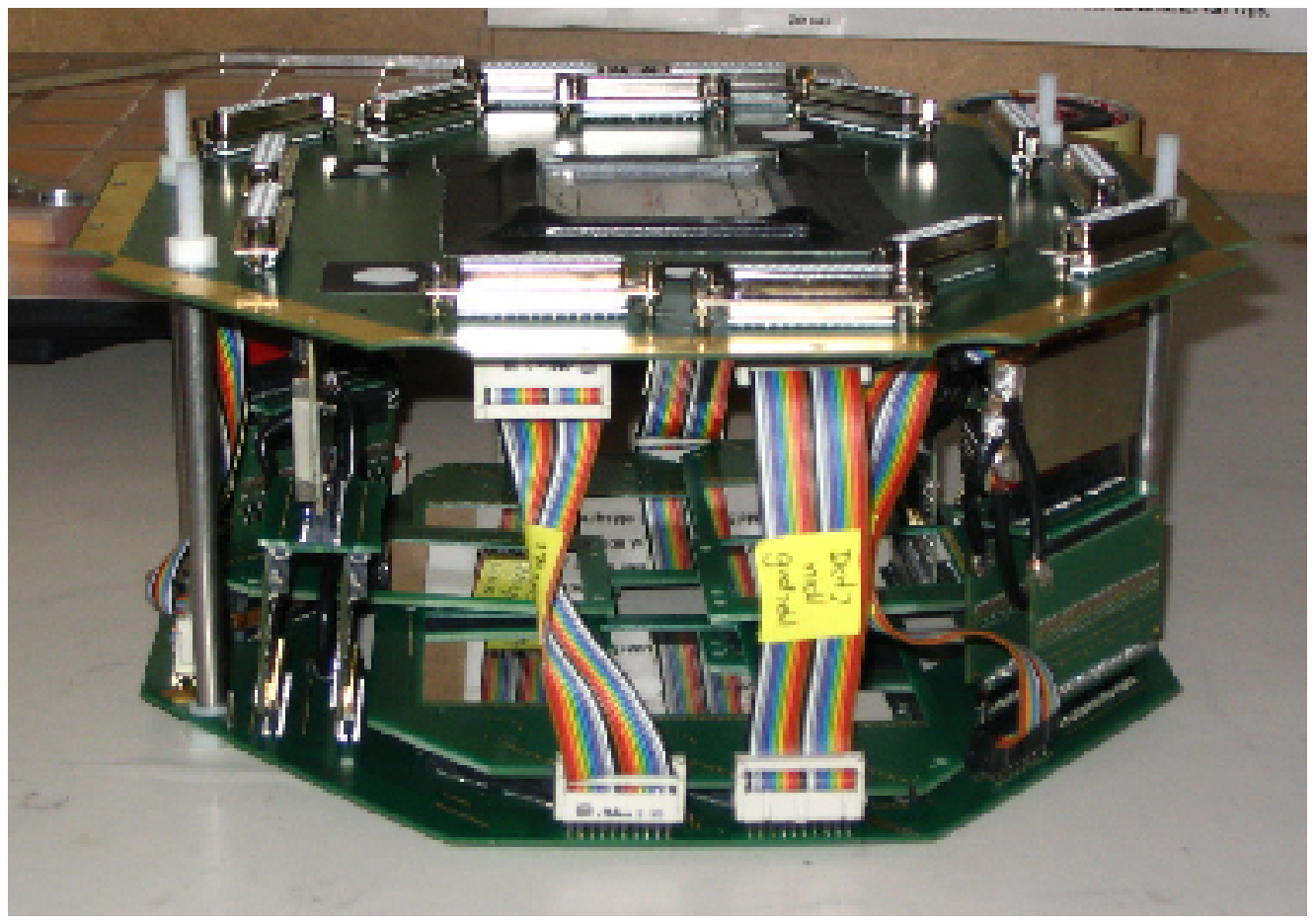}
\end{minipage}
\caption{Left:  Schematic view of the trigger concept. Right:Assembled SI{$\Lambda$}VIO system.}
\label{skiz}
\end{figure}

Recently, the search for deeply bound nuclear states with antikaons has attracted large interest. These states were theoretically predicted  some years ago \cite{aka02} and several experiments exploiting kaon beams have been carried out to verify these hypothesis. Further theoretical efforts have put forward a broad scenario of hypothesis about the binding energy and width of the kaonic bound states are available at the moment \cite{She07}, \cite{Hyo08}. From the experimental point of view some signature have been observed \cite{Iwa03}, \cite{Agn05}  but the limited statistics available so far does not allow to draw any firm conclusion.\\ 
Lately the hypothesis was put forward \cite{yam07}, that the production of the ppK$^-$ might be 
populated quite favorably in the reaction pp$\rightarrow$K$^-$ppK$^+$, due to the large momentum transfer and the small impact parameter of this collision at energies around 3.5GeV.\\
This idea was recently supported by experimental results \cite{yam08} that reports on a significant signature for the ppK$^-$ state for the p+p at 2.85 GeV/c reaction.\\ We have proposed to perform an exclusive measurement using the FOPI detector  exploiting the reaction p+p at 3.5GeV and to build for this purpose a dedicated trigger device.\\
The final state of the pp$\rightarrow$(K$^-$pp)K$^+\rightarrow\Lambda$pK$^+$ reaction involves $\Lambda$ hyperons which can be detected using their decay into p + $\pi^-$ (64\% branching ratio). 
The FOPI \cite{fop95} detector allows the identification of charged particles in an almost 4$\pi$ coverage of the phase space and the reconstruction of secondary emission vertexes  in the region around mid-rapidity.
Moreover, the newly installed RPC detector enhances significantly the identification capability for $K^+$ emitted around mid-rapidity. For this reason the FOPI spectrometer seemed well suited to carry out the above mentioned experiment, provided the development of a trigger concept that can enhance the events we are interested in and that further tracking capability is made available in the forward direction. 
For this reason FOPI  has been extended by a silicon-based $\Lambda$ trigger system, in order to enrich events containing $\Lambda$ candidates and provide an additional position information for particle emitted at high rapidity.\\
The scheme of the $\Lambda$ trigger (SI{$\Lambda$}VIO -- SIlicon $\Lambda$ Vertexing and Identification Online) is shown in the left panel of figure \ref{skiz}. It consists of two detector layers downstream of the target with distances such that the bulk part (about 60\%) of the produced $\Lambda$s decay in between the two layers.
The first layer (SI{$\Lambda$}VIO A) is a single-sided, 1mm thick annular detector segmented in 32 slices, while the second layer  (SI{$\Lambda$}VIO B) consists of a patch-work of 8 rectangular double-sided, 1 mm thick, $40\times 60$mm$^2$ with 1 mm pitch for each side.
The event selection is performed requiring online that the hit multiplicity on the second silicon layer is higher (1 or 2 hits more)  than the hit multiplicity on the first layer. This operation is taken care by the Mesytec analog electronics that reads out the annular and the n side of the 8 rectangular detectors.
The Mesytec shaper provide a trigger signal according to the hit multiplicity on each detector and can be set such to realize the above mentioned trigger condition.
The p-side of the rectangular detectors has been read out with an APV-25 chip which allows a compact readout of all the channels. The assembled detector system is shown in right panel of figure \ref{skiz}, where the boards on which SI{$\Lambda$}VIO A and B are hosted and the APV-25 cards are visible.

\begin{figure}
  \centerline{\includegraphics*[width=9.5cm]{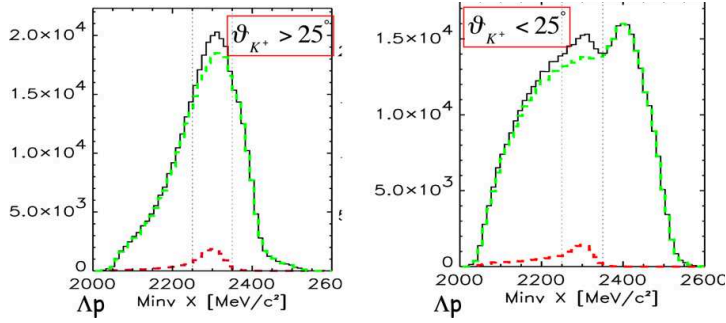}}
  \caption{\label{invM}$\Lambda$-proton invariant mass disitribution corresponding to $K^+$ emitted with large (Left) and small (Right) polar angle (see text for details).}
\end{figure}

A test has been carried out at GSI to test the performance of the trigger system. A proton beam at 3GeV with an intensity of $10^5$ particles/sec has been focused on a plastic target and the full FOPI spectrometer, together with the $\Lambda$-trigger, has recorded data under different trigger condition.
The main trigger (LVL1) has been set requiring at least one charge particle to cross the FOPI-RPC and the FOPI-PLAWA, the time of flight detectors situated at mid- and forward rapidity in the laboratory reference system respectively.
The $\Lambda$ trigger has been set such to accept events with one or more particle hits on SI{$\Lambda$}VIO A in coincidence with two or more hits on SI{$\Lambda$}VIO B.
\begin{figure}[h]
\centering
\includegraphics*[width=6.5cm]{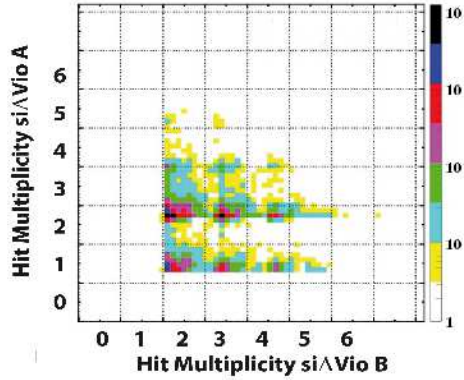}
\caption{Offline particle multiplicity of SI$\Lambda$VIO A versus the particle multiplicity on SI$\Lambda$VIO B.}
\label{mat}
\end{figure}
Figure \ref{mat} shows the particle multiplicity obtained via an offline calibration for events which fullfil the $\Lambda$ trigger condition. One can see how clean the required multiplicity condition is selected by the $\Lambda$ trigger. 
In order to exclude any bias effect introduced by the $\Lambda$ trigger, the phase space distribution of the identified charged particles has been compared for events where only the standard LVL1 trigger was active and for those where a positive decision by the $\Lambda$ trigger was taken. 
\begin{figure}
  \centerline{\includegraphics*[width=6.0cm]{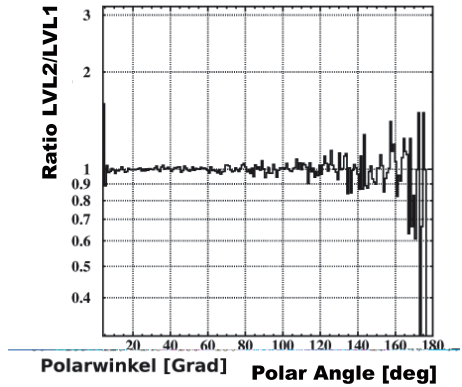}}
  \caption{\label{trigB} Angular distribution of the ratio of charged particles reconstructed in the CDC for $\Lambda$ and LVL1 trigger events.}
\end{figure}
Fig. \ref{trigB} shows the angular distribution of the ratio of the charged particle yield reconstructed in the CDC detector for the two trigger conditions.
One can see that the ratio is rather close to 1 and doesnt show any deformation of the phase space distribution.\\
In overall a reduction of a factor 25 respect to the LVL1 trigger has been obtained applying the $\Lambda$ trigger condition. \\
Additionally, data without the target in place have been collected to verifiy the sensitivita of the trigger system. Out of analysis of these data it was possible to estimate the fraction of 'fake' $\Lambda$ triggers, which amount to about 10\%.\\
The achieved  reduction factor is indeed sufficient to accumulate enough statistics for the measurement of the ppK$^-$ state, assuming a production cross-section of 1$\mu b$ and a beam time period of about 3 weeks \cite{prop06}. Taking these numbers into account, simulations have been carried out to estimate the signal to background ratio expected for the pp$\rightarrow$(K$^-$pp)K$^+\rightarrow\Lambda$pK$^+$ reaction. The ppK$^-$ signal was simulated assuming a mass M=2322MeV/c$^2$ and a width $\Gamma$=61MeV/c$^2$ together with all the backgorund reactions weighted with their cross-section for p+p at 3GeV.\\ 
The details of this analysis are shown in \cite{prop06} and the resulting invariant mass spectrum are shown in fig. \ref{invM}. This shows the $\Lambda$-proton invariant mass spectra in the case the $K^+$ is emitted with a polar angle $\theta >25^{\circ}$ and directly identified by the RPC detector (fig. \ref{invM} left) and in the case the $K^+$ is emitted in the forward direction,  $\theta <25^{\circ}$, and the identification is not done directly but via an event hypothesis.
The lower dashed line depicted in both panels of fig. \ref{invM} shows the reconstructed signal coming from the ppK$^-$ decay. The green curve represents the background which is left after applying all the analysis cuts and which consists mainly of the decay pp$\rightarrow$$\Lambda$pK$^+$. This contribution cannot be ruled out by any of the so far developed cuts and must be subtracted assuming the correct phase distribution. Since the $\Lambda$-proton and $\Lambda-K^+$ final state interaction, which is relevant at lower incident energies \cite{bil98}, should be negligible above 3GeV one can assume that the kinematic is dominated by phase space and hence rather easy to describe. New data by COSY \cite{cos06} deliver us an experimental basis to estimate the contribution of the pp$\rightarrow$$\Lambda$pK$^+$ for the p+p at 3 GeV reaction. On the other hand, theoretical works have already dealt with this item \cite{sib06} and were able to reproduce the available experimental data.\\
The expected signal to background ratio for 3 weeks of measurements amounts to 3/100 and a significance of 5 is estimated for the state candidate.

\section{Summary}
We have described the newly developed SI$\Lambda$VIO trigger system which has been installed in the FOPI spectrometer and shown the feasibility of the proposed measurement. We aim to collect enough statistics to measure a ppK$^-$ state with a production cross-section of about 1 $\mu b$.
Simulations show that we expect a signal to ratio of about 3/100 for the complete statistics.

This work has been supported by the Excellence Cluster 'Universe' of the Technische Universit\"at M\"unchen.

\end{document}